\documentclass{article}%
\usepackage{amssymb}%
\usepackage{amsmath}%
\usepackage{amsfonts}%

\begin{document}

\author{Vladimir K. Petrov\thanks{ E-mail address: vkpetrov@yandex.ru}}
\title{Random membrane model for lattice gluodynamics}
\date{\textit{N. N. Bogolyubov Institute for Theoretical Physics}\\
\textit{\ National Academy of Science of Ukraine}\\
\textit{\ 252143 Kiev, Ukraine. 10.12.2005}}
\maketitle

\begin{abstract}
Model for studying coupling dependence on lattice spacing $a$ in gluodynamics
is suggested. The model predicts $g\rightarrow g_{0}>0$ with $a\rightarrow0$.
Free energy density in the model does not depend on temperature.

\end{abstract}

\section{Introduction}

As it is known, in lattice gluodynamics only the closed surfaces formed by
plaquetts
\begin{equation}
\square_{x\mu\nu}=U_{\mu}\left(  x\right)  U_{\nu}\left(  x+\mu\right)
U_{\mu}^{\dag}\left(  x+\nu\right)  U_{\nu}^{\dag}\left(  x\right)  ;\quad
U_{\mu}\left(  x\right)  \in SU\left(  N\right)
\end{equation}
contribute to the expansion in powers of $\beta$ of the partition function%
\begin{equation}
Z=\int\exp\left\{  \beta\sum_{x}\sum_{\mu,\nu=0}^{3}\square_{x\mu\nu}\right\}
\prod_{x\omega}dU_{\omega}\left(  x\right)  ;\quad\beta\equiv2N/g^{2}%
\end{equation}
where $g$ is the coupling constant. A surface may be self-intersecting, but it
should not intersect another, since in this case those two surfaces must be
treated as one. Since no additional restrictions are imposed, each surface may
be treated as a random membrane.

There is some space in and out of each surface which is unavailable for
another surfaces. We prescribe to any surface $S_{k}$ some effective volume
$V_{k}=V\left(  S_{k}\right)  $, that depends both on area and shape of
$S_{k}$. In fact it is a volume of a minimal imaginary external shell, that
films over each plaquette of the considered surface with the layer of
effective thickness $b$ no less than $1/2$ link. For instance, surface of a
cube with $n$ links on edge, has the volume of $6\times\left(  n+1\right)
^{2}\times1/2 $. One must also take into account some effective redundant
volume $V\left(  S\right)  -bS$, that appears because a gap between the
surfaces can't be reduced to one link distance because their shapes are not
rectangular in general.

Since the states, which differ by of surface shifts give the same
contribution, the number of states contributed by a surface must be
proportional to the volume available for such shifts. Therefore, the pattern
of exactly solvable model \cite{GPZ} may be adjusted to develop a model
intended to compute mentioned closed surface contributions in lattice
gluodynamics. Since within model framework all orders in $\beta$ are taken
into account, model application region may be extended to the weak coupling
area $g<1$.

\section{Model}

Let the four dimensional volume $V_{tot}=N_{\sigma}^{3}N_{\tau}$ contain $r$
surfaces with volumes $V_{k}=V\left(  S_{k}\right)  $, $k=1,...,r$. The
available volume for any surface is $V_{tot}-\sum_{k}^{r}V_{k}$.

Let $\Re_{S}$ be the number of configurations, taken by an isolated shape
$S$ without shifts, then total number of configurations may be computed 'in
the spirit' of a van der Vaals approximation%
\begin{equation}
Z=\sum_{r=0}^{\infty}\sum_{\left\{  S_{j}\right\}  }\frac{\left(  V_{tot}%
-\sum_{k}^{r}V_{k}\right)  _{+}^{r}}{r!}\exp\left\{  t\sum_{k=1}^{r}%
S_{k}\right\}  \prod_{k}^{r}\Re_{S_{k}}\label{Z}%
\end{equation}
where \cite{brych-prud}%
\begin{equation}
x_{+}^{r}=\left\{
\begin{array}
[c]{c}%
x^{r};\qquad x>0\\
0;\qquad x<0
\end{array}
\right.  =\frac{r!}{2\pi i}\int_{+0-i\infty}^{+0+i\infty}e^{px}p^{-1-r}%
dp\label{3}%
\end{equation}
and
\begin{equation}
t=\left\{
\begin{array}
[c]{ccc}%
\ln\left(  g^{-2}/N\right)  & \text{for} & SU\left(  N\right)  ;\quad N>2\\
\ln g^{-2} & \text{for} & SU\left(  2\right)
\end{array}
\right. \label{t}%
\end{equation}
Further we shall change the order of summation and integration. To make it
safe we shift the integration path
\begin{equation}
\frac{\left(  V_{tot}-\sum_{k}^{r}V_{k}\right)  _{+}^{r}}{r!}=\frac{1}{2\pi
i}\int_{c-i\infty}^{c+i\infty}e^{p\left(  V_{tot}-\sum_{k}^{r}V_{k}\right)
}p^{-1-r}dp
\end{equation}
in a such a way that $c>\operatorname{Re}p_{a}$ for singularity $p=p_{a}$ with
the largest $\operatorname{Re}p$ and for the partition function we may write%
\begin{align}
Z  & =\frac{1}{2\pi i}\int_{c-i\infty}^{c+i\infty}e^{pV_{tot}}\sum
_{r=0}^{\infty}p^{-1-r}\sum_{\left\{  S_{j}\right\}  }\prod_{k}^{r}%
\exp\left\{  tS_{k}-pV_{k}\right\}  \Re_{S_{k}}dp\nonumber\\
& =\frac{1}{2\pi i}\int_{c-i\infty}^{c+i\infty}e^{pV_{tot}}\sum_{r=0}^{\infty
}p^{-1-r}\Re^{r}%
\end{align}
where%
\begin{equation}
\Re=\sum_{S=S_{m}}^{\infty}\exp\left\{  tS-pV\left(  S\right)  \right\}
\Re_{S}dS\label{R1}%
\end{equation}
Further we replace summation in $\left(  \ref{R1}\right)  $ by integration
\begin{equation}
\Re=\sum_{S=S_{m}}^{\infty}\exp\left\{  tS-pV\left(  S\right)  \right\}
\Re_{S}dS\rightarrow\int_{S_{m}}^{\infty}\exp\left\{  tS-pV\left(  S\right)
\right\}  \Re_{S}dS\label{R}%
\end{equation}
Possible consequences of such approximation are discussed in Appendix I.

Having collected everything we may finally write%

\begin{equation}
Z=\frac{1}{2\pi i}\int_{c-i\infty}^{c+i\infty}\frac{\exp\left\{
pV_{tot}\right\}  }{p-\Re}dp\label{7}%
\end{equation}

With infinitely increasing lattice volume $V_{tot}=N_{\sigma}^{3}N_{\tau}$
partition function may be with good accuracy written as$\qquad$%
\begin{equation}
Z\simeq\exp\left\{  p_{a}\left(  t\right)  V_{tot}\right\}
\end{equation}
where $p_{a}$ is the singularity of integrand in $\left(  \ref{7}\right)  $
with the largest $\operatorname{Re}p$.

There two possible regimes. The first one is realized when the solution of
\begin{equation}
p-\Re=0,\label{pole}%
\end{equation}
dominates giving the rightmost singularity in a complex $p$-plane. Another
regime is realized when singularity of $\Re$ proper dominates.

Note, that in fact we must include the self volume of $S_{j}$ into the
available one, i.e. replace $\left(  V_{tot}-\sum_{k}^{r}V_{k}\right)  ^{r}$
by $%
{\displaystyle\prod\limits_{j=1}^{r}}
\left(  V_{tot}-\sum_{k}^{r}V_{k}+V_{j}\right)  $. A simple but bulky
computation allows us to write for the partition function%
\begin{equation}
Z=\frac{1}{2\pi i}\int_{c-i\infty}^{c+i\infty}\frac{\exp\left\{
pV_{tot}\right\}  }{p-\Re}\exp\left\{  -\partial\Re/\partial p\right\}  dp
\end{equation}
However, this correction doesn't change the results appreciably, since the
position of singularity of $\partial\Re/\partial p$ is the same as that of
$\Re$.

Domination and possible interchanges of regimes are defined not only by
specific form of $\Re$, but are also determined by the choice of the coupling
constant (and consequently $t$) dependence on lattice spacing $a$. Indeed,
free energy density is%
\begin{equation}
F=-\frac{T}{\mathbf{V}}\ln Z\simeq-\frac{T}{\mathbf{V}}V_{tot}p_{a}\left(
t\right)  =-a^{-4}p_{a}\left(  t\right)
\end{equation}
where $\mathbf{V}=a^{3}N_{\sigma}^{3}$ is a veritable volume of the system and
$T=\left(  N_{\tau}a\right)  ^{-1}$ is the temperature. For any $p_{a}$ which
does not vanish with $a\rightarrow0$ as does $a^{4}\times const$, free energy
density becomes infinite, so such state becomes unavailable for the system.
Therefore, we claim%

\begin{equation}
p_{a}\left(  t\right)  =a^{4}P_{a}\left(  t\right) \label{Scaling}%
\end{equation}
where $P_{a}\left(  t\right)  $ is assumed to be finite. Note that when
dependence $g$ on $a$ is specified and $a\rightarrow0$, free energy density
become constant $F=-\lim_{a\rightarrow0}P_{a}\left(  t\right)  \equiv-P_{0}$.
In particular, it doesn't depend on temperature.

Another two basic assumptions of the model are%
\begin{equation}
\Re_{S}\simeq C\exp\left\{  \mu S\right\}  S^{\gamma-1}\label{Rs}%
\end{equation}
(see Appendix II) and
\begin{equation}
V\left(  S\right)  \simeq bS+\left(  S/\sigma\right)  ^{\alpha}\label{Vs}%
\end{equation}
Some reasons for above assumption\footnote{As it follows from $\left(
\ref{R}\right)  $, role of specific form of $V\left(  s\right)  $ with
$a\rightarrow0$ is essentially diminished due condition $\left(
\ref{Scaling}\right)  $.} are given in Appendix III.

Therefore we may finally write%

\begin{equation}
\Re\equiv\Re\left(  \tau,p\right)  \simeq C\int_{S_{m}}^{\infty}\exp\left\{
-\tau S-p\left(  S/\sigma\right)  ^{\alpha}\right\}  S^{\gamma-1}%
dS\label{R(t,p)}%
\end{equation}
with%
\begin{equation}
\tau\equiv pb-\mu-t
\end{equation}
Integral in $\left(  \ref{R(t,p)}\right)  $ diverges for all $\tau<0$ and
converges for all $\tau>0$. In a singular point $\tau=0$ integral in $\left(
\ref{R(t,p)}\right)  $ converges to finite value
\begin{equation}
\Re\left(  0,p\right)  \simeq\frac{C}{\alpha}p^{-\frac{\gamma}{\alpha}}%
\Gamma\left(  \frac{\gamma}{\alpha},p\left(  S_{m}/\sigma\right)  ^{\alpha
}\right)
\end{equation}
for any $\operatorname{Re}p>0$. If singularity $\tau=0$ of $\Re\left(
\tau,p\right)  $ is leading in a limit $a\rightarrow0$, it specifies the
dependence $t$ on $a$. Indeed, from $p_{a}\left(  t\right)  b=a^{4}%
P_{a}\left(  t\right)  b=\mu+t_{a}$ we get $g^{2}=\allowbreak\frac{1}{N}%
e^{\mu-a^{4}P_{0}b}$ where $P_{0}=\lim_{a\rightarrow0}P_{a}\left(  t\right)
$, hence $g^{2}\rightarrow g_{0}^{2}\allowbreak=e^{\mu}>0$ with $a\rightarrow
0$.

Since $\partial\Re/\partial\tau<0$ for $\tau>0$, function $\Re\left(
\tau,p\right)  $ is monotonously decreasing in this area. Therefore, if the
solution $p_{a}\left(  t\right)  =\Re\left[  p_{a}\left(  t\right)
b-\mu-t,p_{a}\left(  t\right)  \right]  $ of $\left(  \ref{pole}\right)  $
exists, it is unique. There is some range of values of $t$ and $a$ where such solution
exists and varying those parameters we may move $p_{a}\left(  t\right)  $ in
this area. When dependence $t$ on $a$ is specified, i.e. $t=t_{a}$ this area
shrinks into line $p_{a}\left(  t_{a}\right)  $ defined by the single
parameter $a$, and according to $\left(  \ref{Scaling}\right)  $ $p_{a}\left(
t_{a}\right)  \rightarrow0$ with $a\rightarrow0$.

If $t_{a}$ is chosen in a such way that $\lim_{a\rightarrow0}t_{a}\equiv
t_{0}=const$, then $g^{2}\rightarrow g_{0}^{2}\allowbreak=N^{-1}\allowbreak
e^{\mu-t_{0}}>0$. The case of $t_{a}\rightarrow t_{0}\rightarrow-\infty$ is
unacceptable, since it means $g\rightarrow\infty$ with $a\rightarrow0$. It is
easy to check that we cannot choose $g\rightarrow0$ ($t_{a}\rightarrow
t_{0}\rightarrow+\infty$) with $a\rightarrow0$ because there is no solution
$p=\Re\left(  \tau,p\right)  \rightarrow0$ for such choice of $t_{a}$. Indeed,
in this case from $\left(  \ref{R(t,p)}\right)  $ we obtain%

\begin{equation}
\Re\left(  \tau_{a},p_{a}\right)  \simeq C\sum_{n=0}^{\infty}\frac{\left(
-p_{a}\sigma^{-\alpha}\right)  ^{n}}{n!}\int_{S_{m}}^{\infty}\exp\left\{
-\tau_{a}S\right\}  S^{\gamma+\alpha n-1}dS\label{expan I}%
\end{equation}
or%
\begin{equation}
\Re\left(  \tau_{a},p_{a}\right)  \simeq C\tau_{a}^{-\gamma}\sum_{n=0}%
^{\infty}\frac{\Gamma\left(  \gamma+\alpha n,\tau_{a}S_{m}\right)  }{n!}%
\sigma^{-\alpha n}\left(  -\tau_{a}^{-\alpha}p\right)  ^{n}%
\end{equation}
that with \cite{bateman}%
\begin{equation}
\Gamma\left(  \lambda,x\right)  =x^{\lambda-1}e^{-x}\sum_{n=0}^{N-1}%
\frac{\Gamma\left(  1-\lambda+n\right)  }{\Gamma\left(  1-\lambda\right)
}\frac{1}{\left(  -x\right)  ^{-n}}+O\left(  \left\vert x\right\vert
^{-N}\right)
\end{equation}
leads to
\begin{equation}
\Re\left(  \tau_{a},p\right)  \simeq-CS_{m}^{\gamma}\left(  \mu+t_{a}\right)
^{-1}\exp\left(  \left(  \mu+t_{a}\right)  S_{m}\right)
\end{equation}
so $p_{a}=\Re\left(  \tau_{a},p_{a}\right)  \rightarrow-\infty$ with
$a\rightarrow0$, but we claim $p_{a}$ $\rightarrow0$ in such a limit. Hence,
there is no asymptotic freedom in suggested model.

The possibility for QCD (and gluodynamics in particular) to be a
non-asymptotically free theory have been discussed for years. There are
reasons to believe, that QCD is not perturbative at $a\sim0$ \cite{hasenfratz}%
. On the basis of today's numeric computations, it is difficult to anticipate
the behavior of $g$ in the limit of$\ a\rightarrow0$, taking perturbative
calculations as a guidance. Moreover, numerical studies \cite{b-p95} showed
deviations of the Callan-Symanzik $\beta$-function $~\beta_{CS}(g)\ $from
$\ $perturbative result when the correlation length begins to grow. These
deviations are of such a pattern, as if the theory approaches the fixed point
$g_{0}$ at which$~\beta_{CS}=0$ and consequently the theory is not
asymptotically free. Solid arguments in favor of such behavior of $~\beta
_{CS}(g_{0})\ $were given in \cite{PS}. Data on deep inelastic scattering does
not eliminate the fixed point \cite{ss95}. Phenomenological analysis of
available monte-carlo lattice data in the $SU(2)$-gluodynamics shows no contradiction
with the fixed point of $\beta_{CS}(g)$ located at $g_{0}~\simeq0.563$
\cite{bgk}. Analytical estimations \cite{me} also favor $g_{0}\neq0$.

\section{Conclusions}

The pattern of exactly solvable model \cite{GPZ} is adjusted to develop the
random membrane model for lattice gluodynamics. We make use of the fact that
only closed surfaces formed by plaquetts contribute to the expansion of the
lattice gluodynamics partition function in $\beta$ powers. Since within model
framework all orders in $\beta$ are taken into account, model application
region may be extended to the weak coupling area $g<1$. Arguments for the main
assumptions, $\left(  \ref{Rs}\right)  $ and $\left(  \ref{Vs}\right)  $ of
the model are given in Appendix II and Appendix III.

Model predicts $g\rightarrow g_{0}>0$ with $a\rightarrow0$ and independence of
free energy density on temperature.

\section{Appendix I. Discrete Volume}

For the discrete variable $m$ one may define%

\begin{equation}
m_{+}^{r}=\left\{
\begin{array}
[c]{ccc}%
m^{r} & \text{for} & m>0\\
\delta_{0}^{r} & \text{for} & m=0\\
0 & \text{for} & m<0
\end{array}
\right.
\end{equation}
and write a series%
\begin{equation}
\sum_{n=-\infty}^{\infty}n_{+}^{r}e^{-np}=\sum_{n=0}^{\infty}n^{r}e^{-np}%
\end{equation}
which is a discrete version of Laplace transform, called $Z$ - transformation
(see e.g. \cite{doe}). The inverse transform
\begin{equation}
m_{+}^{r}=\frac{1}{2\pi i}\int_{c-i\pi}^{c+i\pi}e^{pm}\sum_{n=0}^{\infty}%
n^{r}e^{-np}dp
\end{equation}
may be used instead of $\left(  \ref{3}\right)  $, so for partition function
we obtain%

\begin{equation}
Z=\sum_{r=0}^{\infty}\frac{1}{r!}\frac{1}{2\pi i}\int_{c-i\pi}^{c+i\pi
}e^{pV_{tot}}\sum_{n=0}^{\infty}n^{r}e^{-np}dp\sum_{\left\{  S_{j}\right\}
}\prod_{k}^{r}e^{tS_{k}-V_{k}\left(  S_{k}\right)  }\Re_{S_{k}}%
\end{equation}
that leads to%
\begin{equation}
Z=\frac{1}{2\pi i}\int_{c-i\pi}^{c+i\pi}e^{pV_{tot}}\sum_{n=0}^{\infty}%
\sum_{r=0}^{\infty}\frac{1}{r!}n^{r}\Re^{r}e^{-np}dp
\end{equation}
where%
\begin{equation}
\Re=\sum_{S}\exp\left\{  pV\left(  S\right)  +tS\right\}  \Re_{S}%
\end{equation}
and one may finally write%

\begin{equation}
Z=\frac{1}{2\pi i}\int_{c-i\pi}^{c+i\pi}\frac{\exp\left\{  pV_{tot}\right\}
}{1-\exp\left\{  \Re\left(  \tau,p\right)  -p\right\}  }dp\label{dscr}%
\end{equation}
Since $\exp\left\{  \Re-p\right\}  $ is entire function of $\Re-p$,
singularities of the integrand in $\left(  \ref{dscr}\right)  $ and in
$\left(  \ref{7}\right)  $ are located at the same position, and the
difference between corresponding expressions for partition functions
disappears with $V_{tot}\rightarrow\infty$.

\section{Appendix II}

For simplicity we consider here the closed two-dimensional surface in the
three-dimensional space, instead of four-dimensional space in suggested model.
The slice between planes $x=x_{c}$ and $x=x_{c}+1$ is a two-dimensional
surface closed in the $x$-direction. Its borders are located in the above
planes. The borders are equal and are closed loops of length $L_{x}$. The area
of this slice is $L_{x}\times1$. For $N$ slices of common area $S_{YZ}$ we get%

\begin{align}
\Re^{\left[  N\right]  }\left(  S_{YZ}\right)   & =\sum_{\left(  L_{x}\right)
}\delta\left(  S_{YZ}-\sum_{x}^{N}L_{x}\right)
{\displaystyle\prod\limits_{x}^{N}}
R_{L_{x}}\nonumber\\
& =\sum_{\left(  L_{x}\right)  }\frac{1}{2\pi i}\int_{c-i\infty}^{c+i\infty
}\exp\left\{  pS_{YZ}\right\}
{\displaystyle\prod\limits_{x}^{N}}
R_{L_{x}}e^{-pL_{x}}dp
\end{align}
or%
\begin{equation}
R^{\left[  N\right]  }\left(  S_{YZ}\right)  =\frac{1}{2\pi i}\int_{c-i\infty
}^{c+i\infty}\exp\left\{  pS_{YZ}\right\}  R^{N}dp
\end{equation}
where%
\begin{equation}
R=\sum_{L}R_{L}e^{-pL}\simeq\int_{L_{\min}}^{\infty}R_{L}e^{-pL}dLdp
\end{equation}

Summing over $N$ we finally get%
\begin{equation}
\Re\left(  S_{YZ}\right)  =\sum_{N=1}^{\infty}R^{\left[  N\right]  }\left(
S_{YZ}\right)  =\frac{1}{2\pi i}\int_{c-i\infty}^{c+i\infty}\frac{\exp\left\{
pS_{YZ}\right\}  }{1-R}dp\label{YZ}%
\end{equation}

\bigskip It is known (see e.g. \cite{Sokal}) that%
\begin{equation}
R_{L}=CL^{\lambda-1}\exp\left\{  mL\right\}
\end{equation}
where $C,\lambda,m$ are constants \footnote{For closed loops without
intersections $C\simeq6/5;\lambda\simeq4/3;m\simeq1$ \cite{Sokal}, but
intersection doesn't change those constants drastically.} and we get%

\begin{align}
R  & \simeq C\int_{L_{\min}}^{\infty}L^{\lambda-1}\exp\left\{  -\left(
p-m\right)  L\right\}  dL\nonumber\\
& =\left(  p-m\right)  ^{-\lambda}C\Gamma\left(  \lambda,\left(  p-m\right)
L_{\min}\right)  \simeq\left(  p-m\right)  ^{-\lambda}C\Gamma\left(
\lambda\right)  .
\end{align}
Having written%

\begin{align}
1-R  & =1-C\Gamma\left(  \lambda\right)  \left(  p-m\right)  ^{-\lambda
}\nonumber\\
& =x\sum_{n=0}^{\infty}\frac{\left(  n+\lambda\right)  !}{\left(  n+1\right)
!\Gamma\left(  \lambda\right)  }\left(  -x\right)  ^{n}\left(  C\Gamma\left(
\lambda\right)  \right)  ^{-\left(  n+1\right)  /\lambda}%
\end{align}
where $x=p-m-\left(  C\Gamma\left(  \lambda\right)  \right)  ^{1/\lambda}$,we
come to an integrand in $\left(  \ref{YZ}\right)  $ having a simple pole in
$p=\mu\equiv m+\left(  C\Gamma\left(  \lambda\right)  \right)  ^{1/\lambda}$.
Since $\lambda>0$ we get $p=\mu>m$, so singular point of $R$ at $p=m$ is
located leftward in complex $p$-plane. Therefore, the pole is a leading
singularity and for $S_{YZ}\gg1$ we may write%

\begin{equation}
\Re\left(  S_{YZ}\right)  \simeq\exp\left\{  \mu S_{YZ}\right\}
\end{equation}

If one assumes that%
\begin{equation}
\Re\left(  S_{X}\right)  \simeq\exp\left\{  \mu S_{X}\right\}
\end{equation}
where $S_{X}=S-S_{YZ}$ part of surface that consists of $x$-planes plaquetts,
one comes to%
\begin{equation}
\Re\left(  S\right)  \simeq\int_{S_{\min}}^{S}\exp\left\{  \mu\left(
S-S_{X}\right)  \right\}  \exp\left\{  \mu S_{X}\right\}  dS_{X}\simeq
S\exp\left\{  \mu S\right\}
\end{equation}
Unfortunately such simple estimation allows to exclude corrections neither in
a power type factor, nor in $\mu$. Nonetheless, even in this case expression
for $\Re\left(  S\right)  $ will not contradict to the assumption $\left(
\ref{Rs}\right)  $.

\section{Appendix III. Packing of surfaces.}

The proper effective volume of the surface $S$ is very close to $bS$, but with
increasing number of surfaces and their areas the packing problem appears.
Despite four century history, this problem has been more or less solved only
for objects of simple form (mainly for spherical ones) \cite{AW,CSZ}. In
particular, for 'random' balls packing in a spherical bag of volume $v$, one
can define the density $\rho=v_{B}/v$ where $v_{B}\ $is the volume of the
balls. The density depends on many conditions, but roughly it may be computed
as \cite{AW}%
\begin{equation}
\rho\simeq\frac{2}{3}-\frac{1}{3}N^{-\frac{1}{3}}\label{rho}%
\end{equation}
where $N$ is the number of balls. One may expect similar density behavior in a
case of random volume of balls. Let $\overline{v}=v_{B}/N$ is an average
volume of a ball, hence one may write%

\begin{equation}
v\simeq\frac{v_{B}}{\frac{2}{3}-\frac{1}{3}\left(  v_{B}/\overline{v}\right)
^{-\frac{1}{3}}}=\frac{3}{2}v_{B}+3\overline{v}^{\frac{1}{3}}v_{B}^{\frac
{2}{3}}+...\label{V-V}%
\end{equation}

If for fixed average volume $\overline{v}$ we increase volume of balls $v_{B}
$, in accordance with $\left(  \ref{V-V}\right)  $, by so doing we increase
$v$. One may expect, that similar relation
\begin{equation}
V\simeq c_{1}V_{B}+c_{2}V_{B}^{\alpha}+....
\end{equation}
is true for a single ball volume $V_{B}$ and volume $V$ which this ball
effectively occupied. Taking this relation as a pattern, we assume that there
exist some constants $b>0;\sigma>0$ and $0<\alpha<1,$ such that for objects of
arbitrary form
\begin{equation}
V\left(  S\right)  \simeq bS+\left(  S/\sigma\right)  ^{\alpha}\label{Vs III}%
\end{equation}

Although expression $\left(  \ref{Vs}\right)  $ is regarded only as the model
assumption, there is at least one more argument in favor $\left(
\ref{Vs}\right)  $. For the convex body the volume of parallel shell with
thickness $b $ is $V\left(  S\right)  =Sb+Mb^{2}+\frac{4\pi}{3}b^{3}$ where
$M$ is an average curvature integral \cite{toth}. Dimensional method allows to
assume that $M\sim S^{1/2}$ that corresponds $\alpha=1/2$ in $\left(
\ref{Vs}\right)  $ and allows to expect that with varying $\alpha$ in $\left(
\ref{Vs III}\right)  $ a reasonable description may be found for a body of
arbitrary form at least for $S\gg1$.

\end{document}